\documentclass[11pt]{amsart}
\usepackage{amsbsy,amssymb,amscd,amsfonts,latexsym,amstext,delarray,
amsmath,graphicx} 

\numberwithin{equation}{section}

\def\C{{\mathbb C}}

\renewcommand{\H}{{\mathbb H}}

\def\cA{{\mathcal A}}

\def\cD{{\mathcal D}}

\def\cH{{\mathcal H}}

\def\Tr{{\rm Tr}}

\def\mass{Y}

\def\fa{{\mathfrak a}}
\def\fb{{\mathfrak b}}
\def\fc{{\mathfrak c}}
\def\fd{{\mathfrak d}}
\def\fe{{\mathfrak e}}

\title[Boundary Conditions of RGE flow in NCG Cosmology]{Boundary Conditions of the RGE flow in the Noncommutative Geometry approach to Particle Physics and Cosmology}
\author{Daniel Kolodrubetz}
\author{Matilde Marcolli}
\address{Physics Department \\
California Institute of Technology \\
Pasadena, CA 91125, USA}
\email{dkolodru@caltech.edu}
\address{Department of Mathematics  \\
California Institute of Technology \\ 
Pasadena, CA 91125, USA}
\email{matilde\@@caltech.edu}
\begin{document}
\maketitle

\begin{abstract}
We investigate the effect of varying boundary conditions on the renormalization group flow in a recently developed noncommutative geometry model of particle physics and cosmology. 
We first show that there is a sensitive dependence on the initial conditions at unification, so
that, varying a parameter even slightly can be shown to have drastic effects on the running 
of the model parameters. We compare the running in the case of the default and the maximal
mixing conditions at unification. We then exhibit explicitly a particular choice of initial
conditions at the unification scale, in the form of modified maximal mixing conditions ,
which have the property that they satisfy all the
geometric constraints imposed by the noncommutative geometry of the model at 
unification, and at the same time, after running them down to lower energies with the
renormalization group flow, they still agree in order of magnitude with the predictions at the electroweak scale.
\end{abstract}

\tableofcontents

\section{Introduction}

The recent work \cite{MaPie} developed cosmological models of the
very early universe based on the particle physics model of \cite{CCM}
derived from noncommutative geometry, via the formalism of spectral
triple and the spectral action functional.  Other results on
cosmological aspects of noncommutative geometry models of particle
physics include \cite{BuFaSa}, \cite{LMMS}, \cite{MaPieTeh}, 
\cite{NeOchSa}, \cite{NeOchSa2}, \cite{Mairi}, \cite{NeSa}. 
In the particle physics model of \cite{Ba}, \cite{CoSM2}, \cite{CCM},
the Lagrangian is obtained by computing the asymptotic expansion at
high energy of the spectral action functional \cite{ChCo} on a noncommutative space
which is the product of an ordinary (commutative) spacetime manifold
and extra dimensions given in the form of a noncommutative space
which is metrically zero-dimensional, but K-theoretically six dimensional.
The choice of the noncommutative space determines the particle
physics content of the model and the gauge symmetries. The 
masses and mixing angles arise geometrically as coordinates on the 
moduli space of Dirac operators of the spectral triple describing
the extra dimensions. In the case of the model developed in \cite{CCM},
the particle physics content is the same as in the $\nu$MSM, namely,
in addition to the particles of the Minimal Standard Model, one has
right handed neutrinos with Majorana mass terms. However, the
model is significantly different from $\nu$MSM when it comes to
the properties of the action functional. In fact, as proved in 
\cite{CCM}, the asymptotic expansion of the spectral action contains
the full Standard Model Lagrangian, with the additional Majorana terms 
for the right handed neutrinos. One has unification of the coupling
constants of the three forces, hence the model has a preferred energy
scale at unification. The asymptotic expansion of the spectral action  
also contains gravitational terms, which are the most interesting part
from the point of view of applications to cosmological models. These
terms contain an Einstein--Hilbert term, a cosmological
term, a conformal gravity term, a non-dynamical topological
term, and a conformal coupling of the Higgs field to gravity.

In the approach to cosmological models developed in \cite{MaPie},
one uses the fact that, at the unification scale, in the terms one
obtains in the asymptotic expansion of the spectral action, the
usual gravitational and cosmological constants
are replaced by effective constants, whose expression at
unification depends upon the Yukawa parameters of the
particle physics content of the model. In \cite{MaPie}, this
is used to derive an early universe model in which one
allows these effective gravitational and cosmological
constant determined by the boundary conditions given 
by the asymptotic expansion of the spectral action, to
run with the RGE flow of the associated particle physics
model, according to the running of the Yukawa parameters
and Majorana mass terms. This allows for a much more
serious variability of the effective gravitational and
cosmological constant in between the unification and
the electroweak epochs of the very early universe (and
in particular during the inflationary epoch) than is usually
considered in other gravity models. In \cite{MaPie}, this
type of running leads to several consequences on early 
universe cosmology, from mechanisms for inflation to effects 
on the gravitational waves and the evaporation law for primordial 
black holes.

For the purpose of the present paper, the specific issues
of the running of the gravitational parameters and of
the resulting interpretations within the model are not
directly relevant, since the results we give here are
specifically about the running with the RGE flow of 
those expressions of Yukawa parameters and Majorana 
masses, which enter the value at unification of the 
asymptotic expansion of the spectral action. 

The analysis performed in \cite{MaPie} depends on the choice of
initial boundary conditions at unification for the renormalization
group flow. The results of \cite{MaPie} are obtained using the default
boundary conditions of \cite{AKLRS}. However, as we show in the
present paper, one obtains significantly different behaviors of the
coefficients of the asymptotic expansion of the spectral action 
by changing boundary conditions. This implies that there will be
the possibility of drawing interesting exclusion curves in the 
space of all possible boundary conditions, on the basis of
comparing the model with cosmological data, for example
through the predictions for the tensor-to-scalar ratio and
the spectral index derived in \cite{MaPie}. 

For the purpose
of the present paper, we first show how one obtains significantly
different curves for the running of the parameters in the action functional
with different choices of the boundary conditions. This shows,
as one would have expected, a sensitive dependence on the
initial conditions at unification, which means that a fine-tuning
problem arises within the model, in the choice of the data at
unification.

The main result of the paper is then to exhibit a
specific choice of boundary conditions, which
we denote {\em modified maximal mixing conditions},
which differ from the default one of \cite{AKLRS}, and which
have the desired properties. Namely, we show that
all the geometric constraints on the data at unification 
derived in \cite{CCM} are satisfied by our choice of
boundary conditions. We also show that, when running
the RGE flow with those boundary conditions, one 
obtains values in the low energy limit that are still
compatible in order of magnitude with the physical 
predictions and observed values at low energy,
as in the case of the default conditions of \cite{AKLRS}.

An important 
aspect of these models is understanding how much nonperturbative
effects in the spectral action may affect the low energy behavior
of the model, since that is the main obstacle to extending to
the more recent universe the cosmological models of \cite{MaPie}.
Our estimates of the low energy behavior when matching 
geometric boundary conditions at unification may also provide
some indirect evidence for the magnitude of such effects.

Recent results of \cite{MaPieTeh} show that, at least in the case
of sufficiently symmetric geometries, the spectral action can be
fully computed non-perturbatively, using the technique of
\cite{CCunc}, and the non-perturbative effects are limited to
the shape of the inflation potential.

\section{The spectral action and the renormalization group flow}

In noncommutative geometry one models the analog of a Riemannian
manifold through the notion of a {\em spectral triple}, consisting of
data $(\cA,\cH,\cD)$ of an involutive algebra, a Hilbert space representation,
and a Dirac operator, which has the compatibility condition of having bounded
commutators with elements of the algebra. Additional structure, in the
form of grading $\gamma$ and real involution $J$ with compatibility
conditions with the data $(\cA,\cH,\cD)$ are also introduced. In the particle
physics context, $\gamma$ corresponds to the two chiralities of fermions and $J$
to the involution that exchanges particles and antiparticles. See \cite{CCM} for
a more detailed account of the underlying mathematical structure, which
we do not recall here. 
The action functional considered in noncommutative geometry models
for particle physics is based on the spectral action \cite{ChCo} for the Dirac operator
of a spectral triple, with additional fermionic terms. In the model of \cite{CCM} this
takes the form
\begin{equation}\label{ActionCCM}
\Tr(f(D_A/\Lambda))  + \frac 12 \langle J \tilde\xi,D_A \tilde\xi\rangle.
\end{equation}
Here $D_A=D+A+\varepsilon' \,J\,A\,J^{-1}$ is the Dirac operator with inner fluctuations given
by the gauge potentials of the form $A=A^\dag=\sum_k a_k[D,b_k]$, for elements $a_k,b_k\in \cA$.
The $\varepsilon'$ is just a function of n mod 8 that gives -1 for n congruent to 1 mod 4 and 1 for all other values
of n. The fermionic term $\langle J \tilde\xi,D_A \tilde\xi\rangle$ should be seen as a pairing of
classical fields $\tilde\xi \in \cH^+=\{ \xi \in \cH\,|\, \gamma \xi =\xi\}$, 
viewed as Grassman variables. For the purpose of cosmological applications,
the most important part of this action functional is the one that comes from the
asymptotc expansion at high energy $\Lambda$ of the spectral action
$\Tr(f(D_A/\Lambda))$, since this contains the gravitational terms and their
coupling to matter.

\subsection{The asymptotic form of the spectral action}

The asymptotic expansion of the spectral action is obtained in the form (see
\cite{ChCo}, \cite{CCM})
\begin{equation}\label{SpActLambda}
\Tr(f(D/\Lambda))\sim \sum_{k\in {\rm DimSp+}} f_{k} \Lambda^k {\int\!\!\!\!\!\!-} |D|^{-k} + f(0) \zeta_D(0)+ o(1),
\end{equation}
where $f_k= \int_0^\infty f(v) v^{k-1} dv$ are the momenta of the function $f$ and the
noncommutative integration is defined in terms of residues of zeta functions 
\begin{equation}\label{zetaD}
 \zeta_{a,D} (s) = \Tr (a \, |D|^{-s}). 
\end{equation}
The sum in \eqref{SpActLambda} is over points in the {\em dimension spectrum} of the
spectral triple, which is a refined notion of dimension for noncommutative spaces, consisting
of the set of poles of the zeta functions \eqref{zetaD}. More explicitly, as proved in \cite{CCM},
when applied to a noncommutative space of the form $X\times F$, with $X$ an ordinary
4-dimensional (Euclidean) spacetime and $F$ the noncommutative space whose algebra
of coordinates is $\C \oplus \H\oplus M_3(\C)$, with $\H$ the algebra of quaternions, the
expansion \eqref{SpActLambda} of $\Tr(f(D_A/\Lambda))$ gives terms of the form
\begin{equation}\label{SAvarchange}
\begin{array}{rl}
S =& \displaystyle{\frac{1}{2\kappa_0^2}  \int\,R
 \, \sqrt g \,d^4 x + \gamma_0 \,\int \,\sqrt g\,d^4 x } \\[3mm]
    +& \displaystyle{ \alpha_0 \int C_{\mu
\nu \rho \sigma} \, C^{\mu \nu \rho \sigma} \sqrt g \,d^4 x + \tau_0 \int R^* R^* \sqrt g \,d^4 x } \\[3mm]
 +& \displaystyle{ \frac{1}{2} \int\,  |D H|^2\, 
\sqrt g \,d^4 x -  \mu_0^2 \int\,  |H|^2\, \sqrt g \,d^4 x }
\\[3mm]
  -&  \displaystyle{ \xi_0 \int\, R \, |H|^2 \, \sqrt g \,d^4 x
+ \lambda_0  \int |H|^4 \, \sqrt g \,d^4 x } \\[3mm]
+& \displaystyle{ \frac{1}{4} \int\,(G_{\mu \nu}^i \, 
G^{\mu \nu i} +  F_{\mu
\nu}^{ \alpha} \, F^{\mu \nu  \alpha}+\, B_{\mu \nu} \, B^{\mu \nu})\, \sqrt g \,d^4 x }.
\end{array}
\end{equation}
The coefficients of these terms are functions 
\begin{equation}\label{coeffsrun}
\begin{array}{ll}
\frac{1}{2\kappa_0^2} = & \displaystyle{\frac{96 f_2 \Lambda^2 - f_0 \fc}{24\pi^2}} \\[3mm]
\gamma_0 = & \displaystyle{ \frac{1}{\pi^2}(48 f_4 \Lambda^4 - f_2 \Lambda^2 \fc 
+\frac{f_0}{4} \fd) } \\[3mm]
\alpha_0 = & \displaystyle{ - \frac{ 3 f_0}{10\pi^2} } \\[3mm]
\tau_0 =& \displaystyle{\frac{11 f_0}{60 \pi^2}} \\[3mm]
\mu_0^2 = &  \displaystyle{ 2 \frac{f_2 \Lambda^2}{f_0} - \frac{\fe}{\fa} } \\[3mm]
\xi_0 = & \frac{1}{12} \\[3mm]
\lambda_0 = & \displaystyle{ \frac{\pi^2 \fb}{2 f_0 \fa^2} } .
\end{array}
\end{equation}
These depend upon the three parameters $f_0$, $f_2$, $f_4$, where $f_0=f(0)$ and for $k>0$
$$ f_k =\int_0^\infty f(v) v^{k-1} dv , $$
where $f_0$ depends upon the common value of the coupling constants at unification energy
and $f_2$ and $f_4$ are free parameters of the model.
The expressions \eqref{coeffsrun} also depend upon the energy scale $\Lambda$ and the
running of these parameters is the main topic of our present investigation. In addition to
the explicit dependence on $\Lambda$ of the coefficients \eqref{coeffsrun} there is also an
additional and very interesting dependence on $\Lambda$ through the coefficients
$\fa$, $\fb$, $\fc$, $\fd$ and $\fe$. These are functions of the Yukawa parameters and 
Majorana masses of the particle physics content of the model, in the form
\begin{equation}\label{abcde}
\begin{array}{rl}
  \fa =& \,\Tr(\mass_{\nu}^\dag \mass_{\nu}+\mass_{e}^\dag \mass_{e}
  +3(\mass_{u}^\dag \mass_{u}+\mass_{d}^\dag \mass_{d})) \\[2mm]
  \fb =& \,\Tr((\mass_{\nu}^\dag \mass_{\nu})^2+(\mass_{e}^\dag \mass_{e})^2+3(\mass_{u}^\dag \mass_{u})^2+3(\mass_{d}^\dag \mass_{d})^2) \\[2mm]
  \fc =& \Tr(M M^\dag)  \\[2mm]
  \fd =& \Tr((M M^\dag)^2)  \\[2mm]
  \fe =& \Tr(M M^\dag \mass_{\nu}^\dag \mass_{\nu}) .
\end{array}
\end{equation}

\subsection{Renormalization group flow}

The particle physics models based on the spectral action functional 
of noncommutative geometry as in \cite{ChCo}, \cite{CCM} are (at present) 
entirely a classical theory. In particular, this means that whenever 
physical predictions are derived in these models using renormalization
group techniques to lower the energy scale from unification, where
the model naturally lives, to ordinary energies, one uses beta
functions and renormalization group equations that are imported
from the ordinary QFT of the specific particle physics Lagrangian
that is obtained from the asymptotic expansion of the spectral
action. This is a delicate issue, since in fact the asymptotic
expansion includes both matter and gravitational terms. The
beta functions and RGE flow adopted here (as in \cite{MaPie}) 
is the one for the extension of the Minimal Standard Model
that includes right handed neutrinos with Majorana mass
terms, while the gravitational effects are not taken into
account in the form of RGE. This is an approximation, since
the non-minimal coupling of the Higgs to gravity in the model
means that one no longer has a clear separation between the
particle and gravitational sectors. Consequences of modified
RGE flows coming from non-minimal couplings to the Higgs
can be found for instance in \cite{BezSha}, \cite{BezSha2}, and in \cite{Oku},
while effects from gravity terms are considered in \cite{Dono}.
For the Minimal Standard Model, there is an extensive literature
on the form of the beta functions and the RGE flow, see for instance
\cite{LX} and references therein. In the case of the noncommutative
geometry model of particle physics of \cite{CoSM}, which did
not yet include right handed neutrinos and Majorana mass terms,
predictions of the Higgs mass were obtained based on using the 
RGE of the Minimal Standard Model.

The RGE analysis of the model of \cite{CCM} considered in
\cite{MaPie}, which we also work with in this paper, differs from the
usual RGE analysis of the Standard Model in the following ways:
\begin{itemize}
\item Instead of the RGE of the Minimal Standard Model, one
considers the equations for the extension with right handed
neutrinos and Majorana masses, as in \cite{AKLRS}.
As in \cite{AKLRS} these are treated by considering
different effective field theories in between the different see-saw
scales (see also \cite{AKLR}, \cite{BLP}). 
\item We vary the initial conditions at unification, by imposing the
geometric constraints derived in \cite{CCM} and at the same
time requiring that the low energy values remain close to the
expected physical values. 
\end{itemize}
The specific information on the NCG model of \cite{CCM} enters here in two
ways: first in selecting the appropriate matter content of the
model (the presence of the extra right handed neutrinos with
Majorana mass terms in addition to the usual Standard Model),
hence the use of the RGE flow of \cite{AKLRS}, and
also in the geometric constraints imposed on the boundary
conditions at unification.

We use, as in \cite{MaPie} the renormalization group equations for the
Standard Model with right handed neutrinos and Majorana mass terms
of \cite{AKLRS}. The numerical results described here are obtained with
a Mathematica code based on the REAP program of \cite{AKLRS} adapted
to our model by the first author.

We recall here that the RGE for this particle physics model is given (at one loop) 
by the beta functions \cite{AKLRS} 
$$ 16 \pi^2 \, \,  \beta_{ g_i } = b_i \, g_i^3  \ \ \  \text{ with }
(b_{SU(3)}, b_{SU(2)}, b_{U(1)}) = ( -7, - \frac{19}{6}, \frac{41}{10}) $$
$$ 16 \pi^2 \, \,  \beta_{\mass_u} =  
\mass_u(\frac{3}{2} \mass_u^\dag \mass_u - \frac{3}{2} \mass_d^\dag 
\mass_d + \fa - 
    \frac{17}{20} g_1^2 - \frac{9}{4} g_2^2 - 8g_3^2 ) $$
$$ 16 \pi^2 \, \,  \beta_{\mass_d} = 
\mass_d (\frac{3}{2} \mass_d^\dag \mass_d - \frac{3}{2} \mass_u^\dag 
\mass_u +  \fa -
\frac{1}{4}g_1^2 - \frac{9}{4}g_2^2 - 8 g_3^2 ) $$
$$ 16 \pi^2 \, \,  \beta_{\mass_{\nu}} =   \mass_{\nu} (
\frac{3}{2}\mass_{\nu}^\dag \mass_{\nu}- \frac{3}{2}
\mass_e^\dag \mass_e + \fa - \frac{9}{20} g_1^2 - \frac{9}{4} g_2^2 ) $$
$$ 16 \pi^2 \, \,  \beta_{\mass_e} = \mass_e (
\frac{3}{2}\mass_e^\dag \mass_e- \frac{3}{2}
\mass_{\nu}^\dag \mass_{\nu}  +\fa  -\frac{9}{4} g_1^2 - \frac{9}{4} g_2^2) $$
$$ 16 \pi^2 \, \,  \beta_{M} = 
\mass_\nu \mass_\nu^\dag M + M (\mass_\nu \mass_\nu^\dag)^T $$
$$ 16 \pi^2 \, \,  \beta_{\lambda} = 6 \lambda^2 - 3\lambda (3 g_2^2 + \frac{3}{5} g_1^2) + 
 3 g_2^4 + \frac{3}{2} (\frac{3}{5} g_1^2 + g_2^2)^2 + 4\lambda \fa -  8 \fb . $$
Notice that we use here the normalization of the coupling constants used in
\cite{AKLRS}, which is different from the one of \cite{CCM}.

In particular, as in \cite{AKLRS}, we solve numerically these equations using
different effective field theories in the intervals of energies between the three
see-saw scales, with matching boundary conditions. Namely, starting from
assigned boundary conditions at unification, one runs the RGE flow down until
the first see-saw scale (the top eigenvalue of the Majorana mass matrix $M$.
Then one integrates out the higher modes by introducing a first effective
theory with Yukawa parameters $Y_\nu^{(3)}$ obtained by removing the 
last row of $Y_\nu$ in the basis where $M$ is diagonal and with Majorana mass
matrix $M^{(3)}$ obtained by removing tha last row and column. One then restarts
the RGE flow for these new variables with matching boundary conditions at the
top see-saw scale, until the second see-saw scale, and so on. One has in this
way effective field theories $(Y_\nu^{(k)},M^{(k)})$, $k=3,2,1$.

We study the effect on this RGE flow of changing boundary conditions at
unification scale, and we then derive consequences for the running of the
coefficients $\fa$, $\fb$, $\fc$, $\fd$, $\fe$ of \eqref{abcde}. 
In the next section we show, as could have been
expected, that the running is highly sensitive to the choice of the intial
conditions at unification. This shows that there is an important fine-tuning
issue in the model related to the assigned values at unification. We then
present in the following section a specific choice of boundary conditions
that meets all the geometric constraints on the model and that produces
realistic values at low energies.

\subsection{A remark on gravitational and Yukawa parameters in the NCG models}

This subsection is not directly relevant to the main result of the paper,
which is simply a statement about the running of the parameters $\fa$, $\fb$, $\fc$, $\fd$, 
$\fe$ of \eqref{abcde}, subject to different choices of boundary conditions 
at unification, with particular attention to those dictated by the geometric
constraints imposed by the model of \cite{CCM} at unification. However,
we include it here to discuss briefly and compare different existing points of view 
on the role of the parameters \eqref{abcde} in the coefficients \eqref{coeffsrun}
of the spectral action expansion.

In the NCG model of \cite{CCM}, the relation \eqref{coeffsrun} between
the coefficients of the asymptotic expansion of the spectral action and
the Yukawa coupling and Majorana mass terms of the particle
physics sector holds only at unification energy. In particular, the
dependence of the effective gravitational and effective cosmological
constants upon the parameters $\fa$, $\fb$, $\fc$, $\fd$, 
$\fe$ of \eqref{abcde} only sets the boundary conditions at unification.
In \cite{CCM} (see also the exposition in Chapter 1 of \cite{CoMa}),
consequently, the running of the gravitational terms of the model
is deduced from the usual approach as in \cite{Dono}, see also
\cite{DouPer},  by which one obtains only a very moderate (or 
essential lack of) running of the gravitational parameters. The running of the particle
physics sector is then ruled, in the NCG models, only by the RGE flow of the matter
Lagrangian, neglecting gravitational effects (with the caveat mentioned
above on the non-minimal coupling with the Higgs).

However, there are cosmological models that include the
possibility of a much more drastic variability of the gravitational
parameters in the very early universe, including in particular
the inflationary epoch. Scenarios with variable  
gravitational constant had been considered early on in  
Jordan--Brans--Dicke gravity, where the
variability happens through the non-minimal coupling of gravity
to a scalar field, and more recently within other modified gravity models, 
and in terms of RGE running \cite{HamWill},
as well as in the context of primordial black holes with
gravitational memory (see for instance \cite{Barrow}, or the recent \cite{Carr} and
references therein). Similarly, a variable cosmological constant
plays a role in various models (see for example \cite{Berman}, \cite{DvaVil}, \cite{YZMa},
\cite{OvCoo}).

In \cite{MaPie}, therefore, a different viewpoint on the effective gravitational
and cosmological constant in the asymptotic expansion of the spectral
action in the NCG models is proposed, and a possible early universe
model is investigated, which only covers the epochs in between the 
unification and the electroweak eras, a period which is expected to
include the inflationary epoch. It is shown that, if one considers an
effective action where the gravitational and cosmological constant
are allowed to run according to the RGE flow of the coefficients
\eqref{abcde} through the expressions \eqref{coeffsrun} and with
the assigned boundary conditions at unification, then one
recovers many of the scenarios predicted by other models with
variable gravitational and cosmological constant, as \cite{Carr},
\cite{dSHW}, \cite{Linde}, and several different mechanism 
for inflation, with predictions about parameters such as the
spectral index and tensor-to-scalar ratio. 

Other recent cosmological applications of \cite{CCM}, such as
those in \cite{BuFaSa}, \cite{MaPieTeh}, \cite{NeOchSa}, \cite{NeOchSa2},
\cite{Mairi}, \cite{NeSa}, follow the more conventional point of view
on the asymptotic expansion of the spectral action and the form of
the coefficients \eqref{coeffsrun}. These different viewpoints
do not directly affect in any way the results of the present
paper, and we only mention them here for the reader's information.

\section{Effects of changing boundary conditions}

The REAP program from \cite{AKLRS} allows the user to adjust the boundary conditions. These 
changes are generally made at $\Lambda_{unif}$, taken here to be $2\times10^{16}$ GeV.
As we understand that only fine tuned initial conditions for the universe allowed its current form,
we expect the boundary conditions at unification energy to drastically effect
the development of our model parameters. We show here, as an example, the different
running of the coefficients $\fa$, $\fb$, $\fc$, $\fd$, $\fe$  of \eqref{abcde} for the
default boundary conditions and for the maximal mixing case. We also show explicitly
the changing behavior of the running of one of these coefficients when one of the 
parameters varies at unification, in order to illustrate the significant dependence
on the initial conditions.

\subsection{The default boundary conditions}

The boundary conditions at unification used in \cite{MaPie} 
are the default boundary conditions of \cite{AKLRS}. These have
the following values.

$$ \lambda(\Lambda_{unif})= \frac{1}{2} $$
$$ \mass_u(\Lambda_{unif})= \left( \begin{array}{ccc}
5.40391\times 10^{-6} & 0 & 0 \\
0 & 0.00156368 & 0 \\
0 & 0 & 0.482902 
\end{array} \right) $$
For $\mass_d(\Lambda_{unif})=(y_{ij})$ they have
$$   \begin{array}{rl}
y_{11}= & 0.0000482105 - 3.382\times 10^{-15} i \\ y_{12}= & 0.000104035 + 
 2.55017\times 10^{-7} i \\  y_{13} =& 0.0000556766 + 6.72508\times 10^{-6} i \\
y_{21}=&  0.000104035 - 2.55017\times 10^{-7} i \\ y_{22}= & 0.000509279 + 
 3.38205\times 10^{-15} i  \\ y_{23}= & 0.00066992 - 4.91159\times 10^{-8} i \\
 y_{31}= & 0.000048644 - 5.87562\times 10^{-6} i \\
 y_{32}= &  0.000585302 + 
 4.29122\times 10^{-8} i \\
 y_{33} = & 0.0159991 - 4.21364\times 10^{-20} i
\end{array}  $$
$$ \mass_e(\Lambda_{unif})=\left( \begin{array}{ccc}
2.83697\times 10^{-6} & 0 & 0 \\
0 & 0.000598755 & 0 \\
0 & 0 & 0.0101789 
\end{array} \right) $$
$$ \mass_{nu}(\Lambda_{unif})= \left( \begin{array}{ccc}
1 & 0 & 0 \\
0 & 0.5 & 0 \\
0 & 0 & 0.1 
\end{array} \right) $$
$$ M(\Lambda_{unif})=\left( \begin{array}{ccc}
-6.01345\times 10^{14} & 3.17771\times 10^{12} & -6.35541\times 10^{11} \\
3.17771\times 10^{12} & -1.16045\times 10^{14} & 5.99027\times 10^{12} \\
-6.35541\times 10^{11} & 5.99027\times 10^{12} & -4.6418\times 10^{12}
\end{array} \right) $$

\begin{center}
\includegraphics[scale=.4]{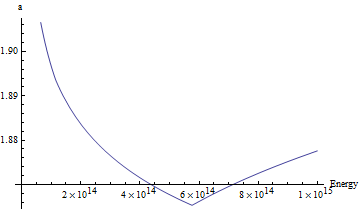}
\includegraphics[scale=.4]{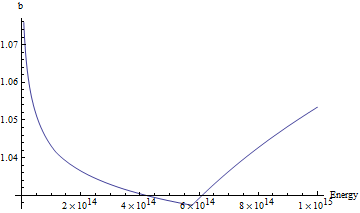}
\includegraphics[scale=.4]{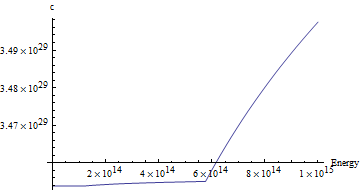}
\includegraphics[scale=.4]{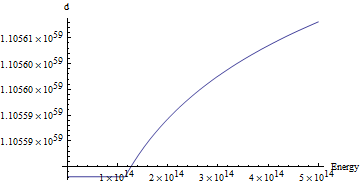}
\includegraphics[scale=.4]{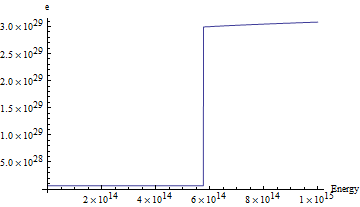}

Figure 1. The running coefficients with default boundary conditions near top see-saw scale
\end{center}

\subsection{Maximal mixing example}

To look at the maximal mixing case, we simply change $\mass_{nu}$ at unification energy. With maximal mixing, our parameters will take these values.

$$\zeta = exp(2\pi i/3)$$
$$ U_{PMNS}(\Lambda_{unif})= \frac{1}{3}\left( \begin{array}{ccc}
1 & \zeta & \zeta^2 \\
\zeta & 1 & \zeta \\
\zeta^2 & \zeta & 1 
\end{array} \right) $$
From the available estimates of the neutrino masses, we get the diagonal mass matrix
$$\delta_{(\uparrow 1)} = \frac{1}{246}\left( \begin{array}{ccc}
12.2 \times 10^{-9} & 0 & 0 \\
0 & 170 \times 10^{-6} & 0 \\
0 & 0 & 15.5 \times 10^{-3} 
\end{array} \right) $$
Finally,
$$\mass_{\nu} = U_{PMNS}^\dagger	 \delta_{(\uparrow 1)}  U_{PMNS}$$

Using this form for $\mass_{\nu}$ and the default boundary conditions on all
the other parameters, we can look at the running coefficients. From the 
figures below, we see that there are vast differences in the 
development of the parameters with this boundary condition change.

\begin{center}
\includegraphics[scale=.4]{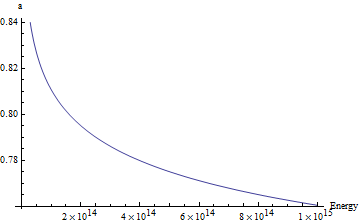}
\includegraphics[scale=.4]{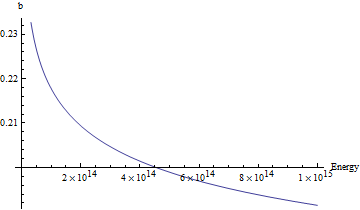}
\includegraphics[scale=.4]{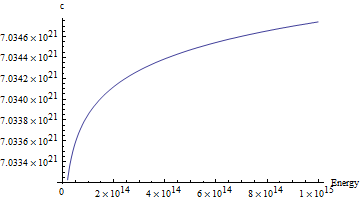}
\includegraphics[scale=.4]{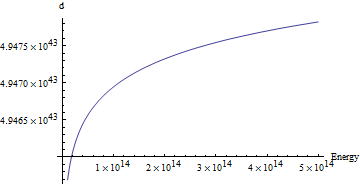}
\includegraphics[scale=.4]{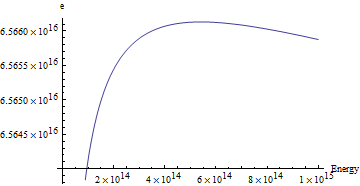}

Figure 2. The running coefficients with maximal mixing boundary conditions near top see-saw scale
\end{center}

\subsection{Running coefficients with changing boundary conditions}

It is possible to get even more interesting behavior by using less standard
boundary conditions. By changing just one parameter we can examine how it
affects the flow of our running parameters. A specific example is the $\mass_{\nu}$
matrix. Using our standard boundary conditions, this matrix is diagonal at
unification energy. We can adjust each of these elements on the diagonal,
which correspond to our neutrino masses, to affect our flow. Using animation
functions in Mathematica, it is possible to get a nearly continuous idea of how
the flow of our parameters develops with our boundary conditions. The figures below
illustrate such a development discretely.

\begin{center}
\includegraphics[scale=.35]{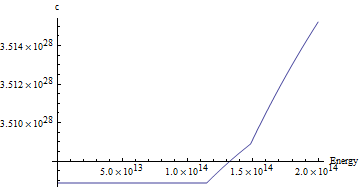}
\includegraphics[scale=.35]{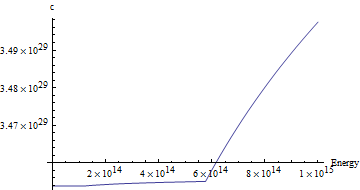}
\includegraphics[scale=.35]{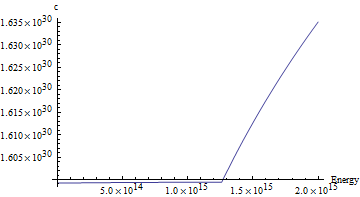}

Figure 2. Coefficient $\fc$ at the upper see-saw scale with the first term of $\mass_{\nu}$ as 0.5, 1.0, and 1.5 respectively
\end{center}

In these diagrams, we notice the transition changing as the upper neutrino mass
varies. The sharp transition at the upper see-saw scale comes from the
program integrating out the heavy neutrino at this scale. The second plot shows
the behavior we expect from the standard conditions. In the first plot
we can see the upper and middle transitions are much closer together than in
our second plot.The final plot shows the transition at a much higher energy,
corresponding to the higher neutrino mass. From these and other such plots,
we learn how the running develops independently by changing different parameters.
Of course, changing multiple parameters complicates this development and is dealt
with in more detail when matching specific boundary conditions.

\section{Geometric constraints at unification}

There are some constraints on the boundary conditions at unification
that are imposed by the underlying geometry of the model. These are derived 
in \cite{CCM}, see also the discussion in \S 1 of \cite{CoMa}. We recall them
here. Not all of these constraints are satisfied by the default boundary
conditions of \cite{AKLRS}, so a first improvement on the model of \cite{MaPie} is to
identify choices of boundary conditions that satisfy these constraints, and
then, among them, eliminate those that produce non-physical predictions. 

We show here how to obtain a choice of boundary conditions that satisfy
all the constraints by modifying the maximal mixing conditions.

\subsection{Constraint on $\lambda$}

A first constraint imposed by the geometry is on the value of the 
Higgs self-coupling $\lambda$ at unification. This satisfies
\begin{equation}\label{lambdaunif}
 \lambda(\Lambda_{unif})=\frac{\pi^2}{2 f_0} \frac{\fb(\Lambda_{unif})}{\fa(\Lambda_{unif})^2} .
\end{equation} 

Looking at our maximal mixing boundary conditions we can calculate that $\lambda(\Lambda_{unif})=2.989$.
By setting it to this value at unification energy in our flow we can ensure that this requirement is met.

\subsection{The $\fa$ parameter and the Higgs vacuum}

The model of \cite{CCM} also relates the parameter $\fa$ to the Higgs vacuum through
the relation
\begin{equation}\label{aHiggs}
 \frac{\sqrt{\fa f_0}}{\pi} = \frac{2 M_W}{g},
\end{equation} 
where $g$ is the common value of the coupling constants at unification and $M_W$ is
the $W$-boson mass. As $M_W$ is directly proportional to $\sqrt{\fa}$, this
condition is a statement of the equality of $f_0$ and the coupling constants
at unification energy.

\subsection{Constraint on $\fc$}

The see-saw mechanism is implemented in \cite{CCM} geometrically, through the
fact that the restriction
of the Dirac operator $D(\mass)$ to the subspace of $\cH_F$ spanned by $\nu_R$,
$\nu_L$, $\bar \nu_R$, $\bar \nu_L$ is of the form
\begin{equation}\label{seesawCCM}
\left( \begin{array}{cccc}
0 & M_{\nu}^\dag & \bar M_R^\dag & 0 \\
M_{\nu} & 0 & 0 & 0 \\
\bar M_R & 0 & 0 & \bar M_{\nu}^\dag \\
0 & 0 & \bar M_\nu & 0 
\end{array}\right),
\end{equation}
where $M_{\nu}$ is the neutrino mass matrix, see Lemma 1.225 of \cite{CoMa}.
This imposes a constraint at unification on the coefficient $\fc$, of the form
\begin{equation}\label{cunif}
 \frac{2 f_2 \Lambda_{unif}^2}{f_0} \leq  \fc(\Lambda_{unif}) \leq \frac{6 f_2 \Lambda_{unif}^2}{f_0} .
\end{equation}
By setting our Majorana mass matrix to 10 times its default value, the inequality can be matched.
In this particular case, the $f_2$ that is used is in the range given in \cite{MaPie}. $f_0$ is
calculated from the coupling constants at unification energy.

\subsection{The mass relation at unification}

Another prediction which is specific to the model of \cite{CCM} is a quadratic
relation between the masses at unification scale, of the form
\begin{equation}\label{massrelation}
\sum_{generations} ( m_\nu^2 + m_e^2 + 3 m_u^2 + 3 m_d^2 ) |_{\Lambda=\Lambda_{unif}} = 8 M_W^2 |_{\Lambda=\Lambda_{unif}} ,
\end{equation}
where $m_\nu$, $m_e$, $m_u$, and $m_d$ are the masses of the leptons
and quarks, that is, the eigenvectors of the matrices $\delta_{\uparrow 1}$, $\delta_{\downarrow 1}$,
$\delta_{\uparrow 3}$ and $\delta_{\downarrow 3}$, respectively, and  
$M_W$ is the W-boson mass. We use the fact that $M_W$ is given as a function of the
model parameters by
\begin{equation}\label{MWdefinition}
\frac{\sqrt{\fa}}{2\sqrt{2}}=M_W.
\end{equation}
So, our equation becomes
\begin{equation}\label{massrelation2}
\sum_{generations} ( m_\nu^2 + m_e^2 + 3 m_u^2 + 3 m_d^2 ) |_{\Lambda=\Lambda_{unif}} =\fa |_{\Lambda=\Lambda_{unif}} .
\end{equation}
In our maximal mixing boundary conditions, we get
\begin{equation}\label{massrelation2}
\sum_{generations} ( m_\nu^2 + m_e^2 + 3 m_u^2 + 3 m_d^2 ) |_{\Lambda=\Lambda_{unif}} =0.6698=\fa |_{\Lambda=\Lambda_{unif}} .
\end{equation}
This value of $\fa$, when converted to conventional units, gives a value of $M_W$ of 72 GeV.
The expected value on $M_W$ is around 80 GeV so these boundary conditions are believable.

\subsection{Modified maximal mixing}
Thus, the conclusion of this analysis is that we obtain a choice of
boundary conditions that satisfies all the geometric constraints
of the geometric model at unification by using our maximal mixing boundary 
conditions as described in the previous section, but with a modified Majorana 
mass matrix and Higgs parameter, as explained here. We refer to the
resulting boundary conditions as the {\em modified maximal mixing conditions}.

We then need to check that, when we run the RGE flow with these
initial conditions at unification, we obtain values at low energies
that are compatible, within order of magnitude, with the 
required physical values. We discus this in the next section.

\section{Low energy physical constraints}

At the electroweak scale, physical data impose other boundary conditions
on some of the Yukawa matrices. Finding the unification scale
conditions that can also match these lower energy requirements is crucial
to the theory.
.
We look at the conditions that are expected from physical data and compare to
the results from the running of the model parameters. We show that our
modified maximal mixing boundary conditions also satisfy the required
constraints at low energy.

\subsection{Boundary conditions at the electroweak scale}
Current predictions at the electroweak scale tell us that the CKM matrix
at $\Lambda_0$ can be taken to be of the form
$$U_{CKM}(\Lambda_0)=\left( \begin{array}{ccc}
0.97419 & 0.2257 & 0.00359 \\
0.2256 & 0.97334 & 0.0415 \\
0.00874 & 0.0407 & 0.999133 
\end{array} \right)$$
Combined with
$$\delta_{(\downarrow3)}(\Lambda_0)=\frac{1}{246}\left( \begin{array}{ccc}
0.00475 & 0 & 0 \\
0 & 1 & 0 \\
0 & 0 & 4.25 
\end{array} \right)$$
we get that the Yukawa parameters for the quarks are given by
$$\mass_{d}=U_{CKM}(\Lambda_0)\delta_{(\downarrow3)}(\Lambda_0)U_{CKM}(\Lambda_0)^\dagger$$
and
$$\mass_{u}(\Lambda_0)=\frac{1}{246}\left( \begin{array}{ccc}
0.0024 & 0 & 0 \\
0 & 1.25 & 0 \\
0 & 0 & 173 
\end{array} \right)$$

Similarly, for the matrix of charged leptons, the known
values and low energy are
$$\mass_e(\Lambda_0)=\frac{1}{246}\left( \begin{array}{ccc}
0.000511 & 0 & 0 \\
0 & 0.1056 & 0 \\
0 & 0 & 1.777 
\end{array} \right)$$

The conditions for the other parameters are all given at the unification
scale.

\subsection{Comparison of expected and measured values}
We use the modified maximal mixing boundary conditions to run the parameters
and compare to the physical boundary conditions at low energy. From this analysis,
we get that the measured Yukawa parameters for quarks are

$$\mass_{d, {\rm measured}}(\Lambda_0)=\frac{1}{246}\left( \begin{array}{ccc}
0.0121 & 0 & 0 \\
0 & 0.128 & 0 \\
0 & 0 & 4.032 
\end{array} \right)$$
and
$$\mass_{u, {\rm measured}}(\Lambda_0)=\frac{1}{246}\left( \begin{array}{ccc}
0.0032 & 0 & 0 \\
0 & 0.9223 & 0 \\
0 & 0 & 248 
\end{array} \right)$$

For the charged leptons, we get the mass matrix
$$\mass_{e, {\rm measured}}(\Lambda_0)=\frac{1}{246}\left( \begin{array}{ccc}
0.000699 & 0 & 0 \\
0 & 0.147 & 0 \\
0 & 0 & 2.51 
\end{array} \right)$$

Comparing these to the expected values at low energies, we see that the order of magnitude
and form of the matrices agree. While the agreement is not exact, it seems that this is the closest
we can get while maintaining the geometric constraints of the model. In order to make the agreement more exact, further fine tuning is required.
\bigskip

\section{Conclusions}

In this paper we investigate the RGE running of the coefficients
$\fa$, $\fb$, $\fc$, $\fd$, $\fe$  of \eqref{abcde}, which appear in
the asymptotic expansion of the spectral action functional of
the noncommutative geometry model of particle physics
of \cite{CCM}. The equations used in the renormalization
group analysis are based on the beta function calculation
of \cite{AKLRS}, for the extension of the Standard Model that
includes right handed neutrinos with Majorana mass terms.

We showed that the running is very sensitive to
the fine tuning of the initial conditions at unification energy. 
We exhibited, as significant examples, the different running
for the default boundary conditions of \cite{AKLRS} and the
maximal mixing conditions, and we also showed the effect
on the running of the coefficients of changing a single 
parameter in the initial conditions at unification.

We then showed that a choice of boundary conditions based
on the maximal mixing, with a modified Majorana 
mass matrix and Higgs parameter at unification, satisfies all
the geometric constraints on the model described in \cite{CCM},
while at the same time gives rise to low energy values that are,
within order of magnitude, in agreement with the expected 
physical values. 

We consider here the asymptotic expansion of the spectral
action in the range of energies from the unification scale
down to the electroweak scale. Within this range of energies,
replacing the non-perturbative form of the spectral action with
its asymptotic expansion is justified, since the error term is
at worse of the order of $\Lambda^{-2}$.  However, it is
known that interesting nonperturbative effects do arise in
the spectral action, as shown in the recent results of
\cite{CCunc}, for example in the form of a slow-roll 
inflation potential. Cosmological implications of these
effects are discussed in \cite{MaPieTeh}. In terms of the
RGE analysis considered here, we find that with our choice
of modified maximal mixing conditions at unification, one
obtains low energy values that are in agreement with
the physical data within order of magnitude, which is
not yet as good an agreement as one could hope for.
This may be an indication that further fine tuning of the
initial conditions may achieve a better fit of the low
energy data, or else that nonperturbative effects may
play a role. This is not completely unlikely, considering
that the nonperturbative effects identified in \cite{CCunc} 
essentially appear in the coupling of Higgs and gravity 
and this in turn can affect the form of the RGE running,
as observed in \cite{BezSha}, \cite{BezSha2}, \cite{Oku}.
These questions will require further investigation.

\bigskip

{\bf Acknowledgment.} This paper is based on the results of Daniel Kolodrubetz's
summer research project supported by Caltech's Summer Undergraduate Research Richter Memorial Fellowship. We thank the referee for many useful comments.

\end{document}